A simple theory is proposed for the dispersive molecular binding of unusually high magnitude due to an enhanced polarizability. Two alternative ways have so far been considered in the literature leading to the polarizability enhancement: (i) a vibronic energy level gap narrowing, as proposed by us with regard to a hypothetical exciton matter, and (ii) a giant electric dipole in a Rydberg state of constituent atoms, as proposed by Gilman with regard to an enigmatic substance building the ball lightning. We now combine the two mechanisms to obtain concrete expressions for the colossal binding energy. The problem is exemplified for a three-level system coupled to the umbrella mode of an ammonia molecule. Other possibilities for the design of enhanced-polarizability molecules are also discussed. The colossal Van der Waals binding is most likely to materialize in hard condensed matter and perhaps less so in soft condensed matter.

1. Introduction

The possibility of colossal Van der Waals (VdW) interactions, ones of unusual strength, occuring in crystalline or amorphous solids as well as in soft condensed matter has lately been on debate in the literature [1-3]. There now seems to be a consensus that colossal dispersive interactions may arise from the enhanced polarizability of specific systems.

Examples have been considered of the polarizability of vibronic excitons enhanced by interlevel gap narrowing (Holstein effects) in crystalline solids and in molecular systems [1,3], or of the polarizability of molecules with constituent atoms excited to a Rydberg state characterized by a high electric dipole moment.[2] The colossal VdW coupling has been suggested to build up the cohesion of a ball lightning characterized by both a strong binding and a low shear modulus [2,3].

Still, the colossal VdW binding has only been inferred indirectly from experimental observations suggesting that molecular systems may exhibit unusually strong a cohesion [4]. While the general qualitative features of the colossal dispersive coupling have already been described in the foregoing articles, we feel the necessity of summarizing the basic results and propose illuminating examples for an experimental verification.

This paper is organized as follows: We begin with summarizing the basic VdW equations known obtainable to second-order perturbation for a system with dipolar interactions [5-7]. We subsequently consider two distinct cases of polarizability enhancement by first introducing the vibronic coupling leading to energy level gap narrowing and then the Rydberg state benchmarked by a high electric dipole. These two ways of elevating the polarizability may work separately bringing about a huge enhancement but if they interfere the effect multiplies. After this general theoretical background, we consider a three-level system composed of (i) a bound ground state, followed by (ii) a Rydberg state, and (iii) another bound excited state in the increasing energy order. All the three states mix vibronically by an appropriate vibrational mode, the (i) - (iii) and (ii) - (iii) pairs mixing strongly, while the (i) - (ii) pair mixing only weakly. The energy level system resembles the levels of an ammonia molecule with the $A_2''$ umbrella mode as the mixing vibration. We carry out numerical calculations using the parameters pertinent for the $NH_3$ molecule [8]. This molecule is interesting in that its adiabatic potentials display double-well characters in both the ground state and the excited Rydberg state which makes it possible to study the two polarizability-enhancing agents outright for the same molecular system. This analysis will be expected to draw the attention of experimentalists to verify the predicted large magnitude of the dispersive interactions in the $NH_3$ molecular ensemble.

## 2. Polarizability enhancement

The electrostatic polarizability $\alpha$ of an entity within a crystalline environment is usually regarded as spatially confined to the characteristic unit cell volume. As a matter of fact, the ultimate value of the electric dipole moment induced by a Coulomb field within the unit cell is obtained from $p \sim ed \sim \alpha \, (e / \kappa d^2)$ wherefrom it follows $\alpha \propto \kappa d^3$ where $d$ is the lattice spacing. The implication is that the electrostatic polarizability should not exceed the unit cell volume. This conclusion drawn for a simple Coulomb potential may not be holding true for more complex situations, e.g. for screened Coulomb potentials.

The quantum mechanical analysis relates the polarizability $\alpha \sim p^2 / \Delta\varepsilon$ to the squared transition dipole $p \sim ed$ and to the energy difference $\Delta\varepsilon$ associated with the electronic transition at a polarizable unit in both hard and soft condensed matter. The above relationship opens at least two ways of controlling the polarizability through varying the numerator or the denominator, respectively. As a result, there are a few known cases in which the polarizability may greatly be enhanced over its dimension-limited coulombic maximum:

(i) Photoexciting an atom to a Rydberg state elevates vastly its associated transition dipole moment $p$. Indeed, for a polarizable Rydberg atom $d$ should rather be meant to be that atom's radial dimension which exceeds grossly the lattice parameter. Consequently, the polarizability of an excited Rydberg atom may be largely superior to the polarizability of that same atom in a nonexcited medium.

(ii) For a two-level system (TLS), another source of enhancement derives from the polarizability being inversely proportional to the interlevel gap energy $\Delta\varepsilon$. Here because of the gap reduction by virtue of Holstein's polaron effect, $\Delta\varepsilon$ may largely drop relative to its net electronic value due to the system coupling to an appropriate vibrational mode.

## 3. Theory of the Van-der-Waals binding

### 3.1. Uncoupled TLS

The theory of 'normal-strength' VdW interactions in atomic systems has been described in detail elsewhere [7]. Here we shall reproduce some of the basic results related to the binding of atomic-like species in $s$-state. For our specific purpose we also assume these atoms to have only one excited state in addition to the ground state which is a kind of TLS. In fact, TLS have been found to play a basic role in the physical properties of glasses, though they have been inferred for crystalline materials as well [9].

To begin with, we first consider TLS to be vibrationally decoupled, so as to reveal their purely electronic behavior. This restriction will be lifted later to see just how the VdW binding is affected by the vibronic coupling.

The dynamic polarizability of a TLS in ground state will read then

$$\alpha(\omega) = (e^2 f_{0n}/m) [(\omega_{n0}^2 - \omega^2) - i\eta \times \text{sign}\omega]^{-1} , \quad \eta \to 0 \tag{1}$$

where

$$f_{0n} = (2m/\hbar^2) \Delta\varepsilon |\Sigma_i < z_i >|^2$$

is the oscillator strength of the dipolar transition to the first excited state $n$, $z_i$ is the coordinate of the i-th electron (most often a single electron will be implied), $\Delta\varepsilon$ is the interlevel energy gap, $\omega_{n0} = \Delta\varepsilon/\hbar$. The binding energy of two TLS separated at $R$ large with respect to their linear dimensions (when retardation effects can be discarded) is

$$U(R) = (3\hbar/\pi R^6) \int_0^\infty d\omega \, \alpha_1(i\omega) \alpha_2(i\omega) \tag{2}$$

Inserting (1) into (2) and integrating yields

$$U_{\text{VdW}}(R) = (6e^4/R^6) |\Sigma_i < z_i^{(1)} >|^2 |\Sigma_i < z_i^{(2)} >|^2 / (\Delta\varepsilon_1 + \Delta\varepsilon_2)$$

For identical TLS $\Delta\varepsilon_1 = \Delta\varepsilon_2 = \Delta\varepsilon$ and introducing the transition dipoles $U_{\text{VdW}}(R)$ turns in

$$U_{\text{VdW}}(R) = (3/R^6) |<p^{(1)}>|^2 |<p^{(2)}>|^2 / \Delta\varepsilon \tag{3}$$

where $p^{(1)} = p^{(2)}$ under the above-mentioned assumptions. Finally, introducing the static polarizability from (1) at $\omega = 0$:

$$\alpha(0) = (e^2 f_{0n} \hbar^2/m) / (\Delta\varepsilon)^2 = 2p^2/\Delta\varepsilon \tag{4}$$

we also obtain the equivalent form

$$U_{\text{VdW}}(R) = \tfrac{3}{4} \Delta\varepsilon [\alpha^{(1)} \alpha^{(2)} / R^3]^2 \tag{5}$$

At this point we note that the quantity ½ $\omega_{n0}$ = ½ $\Delta\varepsilon/\hbar$ signifies the TLS interchange frequency in tight-binding theory.

### 3.2. Vibronically coupled TLS

We next relax the decoupling condition to switch on the interaction of TLS with an asymmetric mode, an odd vibrational mode if the TLS ground and excited states are of the opposite parities, e.g.

$A_{1g}$ (even parity) and $T_{1u}$ (odd parity), respectively, for a cubic lattice ($O_h$ symmetry group). A TLS is said to couple strongly to an asymmetric mode if its coupling energy $\varepsilon_{JT}$ exceeds a quarter of the interlevel gap energy, $\varepsilon_{JT} > \frac{1}{4} \Delta\varepsilon$. Otherwise the coupling is termed weak ($\varepsilon_{JT} < \frac{1}{4} \Delta\varepsilon$). In the latter case the adiabatic potentials associated with TLS are single-well anharmonic parabolas centered at the origin (the symmetric configuration), as in Figure 1 (middle pair). In the former, the lower state parabola turns double well peaking at the symmetric configuration with two lateral minimums bottoming at two broken-symmetry configurations corresponding to left- or right- handedness, Figure1 (extreme pair). In particular, Figure1 (middle pair) depicts the mode-softening situation when $\varepsilon_{JT} = \frac{1}{4} \Delta\varepsilon$. The system performs tunneling transitions between the different-handed configurations restoring on the average the original high symmetry. This justifies applying the present *s*-state formulas to describing lower-symmetry situations. Although adopting a 1D model to present our arguments here, multimode extensions can be made straightforwardly.

When the coupling is strong TLS goes spontaneously from higher symmetry to lower symmetry and its energy is lowered. An electric (vibronic) dipole appears in broken symmetry, the left-hand and right-hand dipoles being equal in magnitude and opposite in sign. We stress that the dipole formation is only possible on TLS coupling to an inversion-symmetry breaking mode. The magnitude of the lateral-well vibronic dipole has been found to be [10]

$$p_{vib} = p \sqrt{[1 - (\Delta\varepsilon / 4\varepsilon_{JT})^2]} \leq p \tag{6}$$

wherefrom it is seen to vanish at the mode-softening point and to increase in magnitude towards its ultimate value *p* as the coupling energy is increased. Another quantity undergoing change on coupling is the interlevel energy gap $\Delta\varepsilon$. The well-known vibronic effect (Holstein's reduction) on the gap energy is the gap narrowing:

$$\Delta\varepsilon_{vib} = \Delta\varepsilon \times \exp(-2\varepsilon_{JT} / \hbar\omega) \tag{7}$$

Due to the exponential to the right the gap reduction can be quite significant. The coupled vibrational frequency is also normalized according to

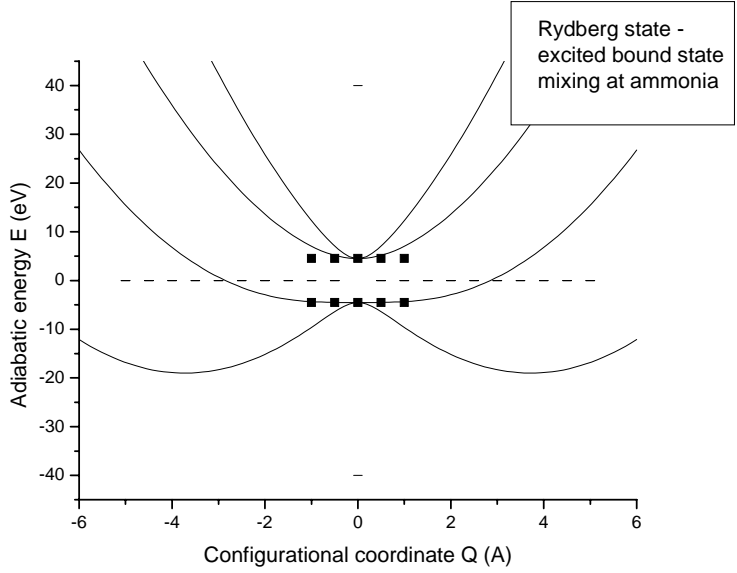

Figure 1

Adiabatic potentials from Eq. (12) for mixing the N atom Rydberg state (lower) with the excited bonding state (upper) at the $NH_3$ molecule. Static states are marked by filled squares, vibronically-couples states are depicted by solid lines. The intermediate potential pair is at the mode-softening point with $G = 3.467$ eV/Å, the extreme lowermost – uppermost pair is in the small-polaron regime for $G = 10$ eV/Å. The remaining parameters: mode stiffness $K = 2.67$ eV/Å$^2$, interlevel gap energy $\Delta\varepsilon = 9$ eV.

$$\omega_{ren} = \omega \sqrt{[1 - (\Delta\varepsilon/4\varepsilon_{JT})^2]} \tag{8}$$

The vibronic binding energy will now read

$$U_{VdWvib}(R) = (3/R^6) |<p_{vib}^{(1)}>|^2 |<p_{vib}^{(2)}>|^2 / \Delta\varepsilon_{vib} \tag{9}$$

We see that all basic arguments regarding the decoupled TLS will hold good for the vibronically-coupled TLS, provided the interlevel gap $\Delta\varepsilon$, the vibrational frequency $\omega$, and the electric dipole $p$ are replaced by their respective counterparts $\Delta\varepsilon_{vib}$, $\omega_{ren}$, and $p_{vib}$, respectively. The emerging vibronic picture may be called "renormalized". Due to the reduced energy gap $\Delta\varepsilon_{vib}$ in the denominator, Eq. (9) predicts a net enhancement of the binding energy over its decoupled value in Eq. (3). The ratio of renormalized to decoupled binding can be found easily and amounts to

$$U_{VdWvib}(R) / U_{VdW}(R) = [1 - (\Delta\varepsilon/4\varepsilon_{JT})^2]^2 \exp(2\varepsilon_{JT}/\hbar\omega) \tag{10}$$

The binding profit due to vibronic coupling is seen to decrease steadily as the mode-softening point ($4\varepsilon_{JT} \sim \Delta\varepsilon$) is approached from above. Towards the opposite extreme at the small-polaron limit ($4\varepsilon_{JT} \gg \Delta\varepsilon$), setting $2\varepsilon_{JT}/\hbar\omega \sim 10$ as a typical estimate, we get $U_{VdWvib}(R)/U_{VdW}(R) \sim 2\times 10^4$ to illustrate the colossal character of the vibronic enhancement. It will be illustrated by calculations in Section 5.1 that the interlevel gap reduction due to the vibronic coupling and, therefore, the vibronic enhancement is not a solely collective (polaron) effect [11], but is also inherent to the isolated molecule.

Subject to our particular attention, the static vibronic polarizability

$$\alpha_{vib}(0) = 2\, p_{vib}^2 / \Delta\varepsilon_{vib} \tag{11}$$

will be elevated as the coupling energy is increased, both the increasing vibronic dipole in the numerator and the decreasing vibronic gap in the denominator acting to raise the polarizability.

## 4. Rydberg states

The Rydberg states are large-orbit hydrogen-like states (the order of 1 cm of a radius!) of optical electrons in atoms with large azimuthal quantum numbers [12]. These states should be reached on excitation within the UV range. The associated transition dipole $p$ will therefore be very large and so will the related static vibronic polarizability. Consequently, a net enhancement of the VdW binding energy can be expected to result from the excitation amounting to several orders of magnitude over the intermolecular value between *1* to *10* meV. The Rydberg state is expected to be short-lived though, so that the peak yield of Rydberg density $\propto I\tau$ (*I*-light intensity, $\tau$-lifetime) may not be high.

The VdW binding energy (5) of the uncoupled TLS, moderate under normal conditions, ia liable to colossal intensification on photoexcitation to a higher lying Rydberg state which brings about the elevation of the electrostatic dipole $p$. The colossal effect of the electrostatic dipole is quadratic in the static polarizability (4) and quartic in the binding energy (5).

The phonon-coupled TLS exhibiting an enhanced polarizability and colossal VdW binding is simultaneously liable to further intensification through electrostatic dipole elevation upon photoexcitation to a higher-lying Rydberg state. We stress that the two effects are independent of each other and probably act in the additive manner with no interference

## 5. Example: The ammonia gas molecule

We shall first consider in some detail the case of a molecule that may happen to play a crucial role in our understanding of the colossal VdW binding. As stated above, ammonia is a textbook example of broken-inversion-symmetry molecular cluster. Ammonia's is the case of a $D_{3h}$ symmetry $NH_3$ cluster in the unstable planar configuration in which the N atom resides in the center of an equilateral triangle formed by the three H atoms. The molecular structure is stabilized by the nitrogen moving out of the $H_3$ plane to form a triangular pyramid with N residing at its vertex. Moving N out of the $H_3$ plane breaks the inversion symmetry of the planar cluster. This is an example of a $D_{3h} \rightarrow C_{3v}$ symmetry conversion. The asymmetric $NH_3$ cluster has two distinct equilibrium configurations, left-handed and right-handed, with corresponding inversion dipoles equal in magnitude but opposite in sign. At ambient temperature the system makes flip-flop jumps from one configuration to the other one as nitrogen passes easily through the hydrogen plane reverting the molecule. Because of these jumps the resultant electric dipole smears. Nevertheless, a system of ammonia molecules is polarizable electrically which gives rise to the VdW intermolecular coupling. As we have seen in Section 3.2, the vibronic mixing may build up a great deal of polarizability for a colossal VdW binding.

## 5.1. Colossal Van-der-Waals binding in ammonia gas

Literature calculations of adiabatic potentials of $NH_3$ molecules show that the instability of the inversion-symmetric planar configuration in $^1A_1'$ ground-state is lifted through the mixing of $^1A_1'$ with the second higher-lying $^1A_2''$ excited state by the $A_2''$ mode coordinate of the $D_{3h}$ point group [8]. From the data tabulated therein we get $\varepsilon_{JT} = 0.16$ eV at $\Delta\varepsilon = 5$ eV (first excited state) and $\varepsilon_{JT} = 5.16$ eV at $\Delta\varepsilon = 14$ eV (second excited state) relative to the mixing with the ground state. We see that the small-polaron criterion $4\varepsilon_{JT} / \Delta\varepsilon \gg 1$ is met, though only approximately, for the ground-state to second-excited-state mixing which lifts the planar instability in the ground state. The stiffness in any state is $K = 0.427 \times 10^5$ dyn/cm $= 2.67$ eV/Å$^2$. In so far as the $A_1'$ and $A_2''$ modes are in-phase hydrogen vibrations along with an counter phase nitrogen vibration [8], the common oscillator mass is $M = [3/M_H + 1/M_N]^{-1} \sim 1/3$ and we arrive at a bare mode frequency of $\omega = 2.78 \times 10^{14}$ s$^{-1}$ $= 0.18$ eV. Using these data we calculate a Holstein reduction factor of $\exp(-2\varepsilon_{JT}/\hbar\omega) = 1.26 \times 10^{-25}$ and $\Delta\varepsilon = 1.76 \times 10^{-24}$ eV. Setting $p_{12} = 1$ eÅ we get a huge vibronic polarizability of $\alpha = 6.52 \times 10^{-12}$ cm$^3$ $= 6.52 \times 10^{12}$ Å$^3$. Using that estimate we calculate $U_{VdWvib} = 76$ meV at $R_{ij} = 2$ Å for $\kappa = 1.5$. This is more than *30* times higher than the exciton binding energy in Ge! The ammonia adiabatic potentials along the $A_2''$ mode coordinate are shown in Figure 2, as calculated from the literature data [8]. These potentials are computed using the equations:

$$\varepsilon_{AD}(Q) = \tfrac{1}{2}\left(KQ^2 \pm \sqrt{(4G^2 + \Delta\varepsilon^2)}\right) \qquad (12)$$

with parameters as follows: $G = 5.24$ eV/Å (electron-mode coupling constant), $K = 2.67$ eV/Å$^2$ (stiffness), $\Delta\varepsilon = 14$ eV (gap energy). $Q$ is the coupled mode coordinate.

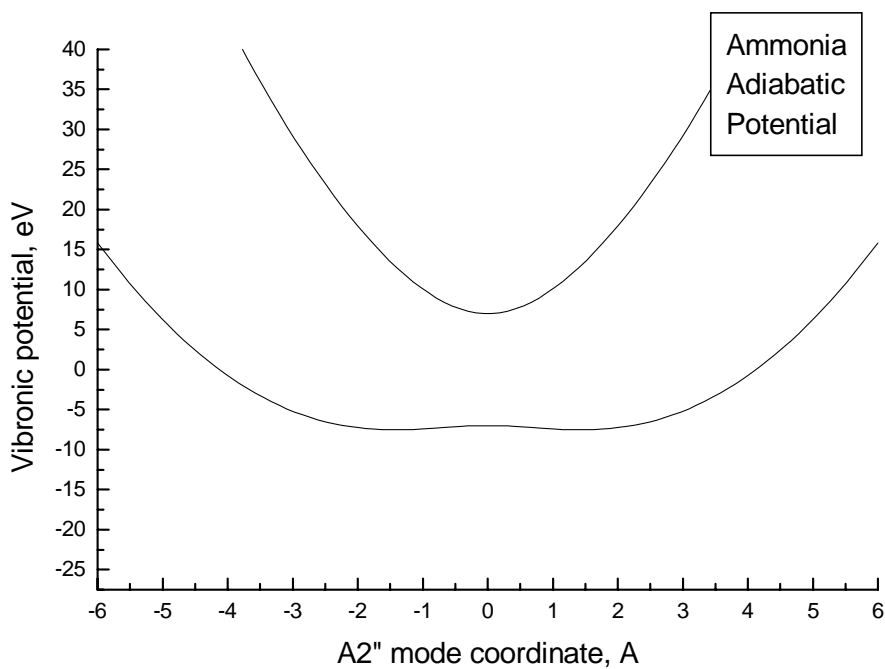

Figure 2

Ammonia adiabatic potentials along the $A_2''$ mode coordinate ($D_{3h}$ point group) rendering unstable the planar configuration of the $NH_3$ molecule. Atomic displacements in the umbrella mode: $H\downarrow$-$H\downarrow$-$H\downarrow$-$N\uparrow$. The instability is lifted as the N atom is pushed stably out of the $H_3$ plane to reside at the vertex of a regular $NH_3$ pyramid. The two pyramidal configurations, left-handed and right-handed, pertain to the two lateral minima of the ground-state potential. The position of a lower-lying excited state in which the N atom is in a Rydberg state is intermediate.

Further, it will be interesting to see whether a strongly coupled vibronic phase can also appear in an excited state of the ammonia molecule. This may account for the occurrence of a colossal VdW binding in the atmosphere under photoexcitation by a linear lightning [3]. For a vibronic exciton phase to form, the stringent requirement is that the molecular cluster should possess a double-well dependence on a mixing mode coordinate in an *excited electronic state*. This is not the case of the ammonia molecule which has a double-well *ground electronic state* along the $A_1'$ mode coordinate, as well as two single-well $A_2''$ excited states [8,13,14]. In other words, the higher symmetry planar $NH_3$ configuration is unstable in ground state and stable in the excited states. Consequently, ammonia may build up a strongly coupled phase only in ground state, while its excitonic states will rather be weakly coupled. Accordingly, ammonia may only build up dark VdW cohesive clouds.

To seek a luminous phase, we consider the vibronic coupling at the N atom excited Rydberg state, whose position is mentioned in the caption to Figure 2. Calculations have shown this state not to mix strongly with the ground state of the ammonia molecule [8]. Instead, the next higher lying excited state has been found to mix strongly with the ground state. We see that the Rydberg state not contributing to the chemical bond at the $NH_3$ molecule, it should mix with the next higher-lying bonding state along the umbrella mode coordinate, for otherwise the system should disintegrate once in a Rydberg state. Strong mixing with the bonding state will produce a double-well character for the adiabatic potential of the Rydberg state. Indeed, in as much as this state is only 9 eV below the next excited bonding state, it will take a coupling constant $G > 3.47$ eV/Å to produce the strong mixing. We therefore appeal for improved numerical calculations of the $NH_3$ potentials. For illustrative purposes we have depicted the $1^{st}$ - $2^{nd}$ excited state mixing by the $A_2''$ mode at the ammonia molecule in Figure 1. The parameters used are $G = 10$ eV/Å, $K = 2.67$ eV/Å$^2$, $\Delta\varepsilon = 9$ eV leading to $\varepsilon_{JT} = 18.73$ eV (the Jahn-Teller energy) and $\varepsilon_B = 14.50$ eV (the interwell barrier). We note that our adopted $G$ value in Figure1 is much too arbitrary to guarantee drawing any decisive conclusions.

### 6. Another example: The methane gas molecule

The vibronic origin of the dynamic instability of the molecular systems considered hereto was given a more complete and rigorous treatment. The nonvibronic contribution to the curvature of the adiabatic potential, due to nuclear displacements under a fixed electronic density distribution, was always positive, and hence the only reason for a dynamic instability was the pseudo Jahn-Teller effect. For some examples of special interest (planar equilateral $NH_3$, planar square $CH_4$, and linear $H_3^+$), the molecular excited states, responsible for the instability of the ground state, were shown by means of *ab-initio* calculations.

Similar adiabatic potentials as in Figure 2 have been obtained using Eq. (12) for $CH_4$ ($G = 4.36$ eV/Å, $K = 0.29$ eV/Å$^2$, $\Delta\varepsilon = 11$ eV) [8,15]. In this case again the dipolar double-well instability has proved inherent to the ground electronic state, while the excited electronic state has remained single-well.

The adiabatic potentials of the $CH_4$ methane molecule (symmetry $D_{4h}$) are shown in Figure 3 against the coordinate $Q$ of the propeller $B_{2u}$ vibrational mode [8,15]. The vibronic coupling effects a $D_{4h} \rightarrow C_{4v}$ symmetry conversion in which the four H atoms form a planar quadrate as the C atom goes out of the plane from its position at the center of the quadrate. From the above parameters we get $\varepsilon_{JT} = 32.77$ eV and $\varepsilon_B = 27.51$ eV These may be compared with the corresponding values for the $NH_3$ ammonia molecule: $\varepsilon_{JT} = 5.16$ eV and $\varepsilon_B = 0.53$ eV. We see that while the ammonia barrier is low allowing for interweel transitions at ambient temperature, the methane barrier is much too high to allow for any sizeable interwell transition rate. Consequently, the methane molecule remains practically frozen in one of the lateral wells which renders a permanent electric dipole moment to it. As a result, methane molecules may bind together by means of permanent dipole-dipole coupling, to be distinguished from the VdW interaction.

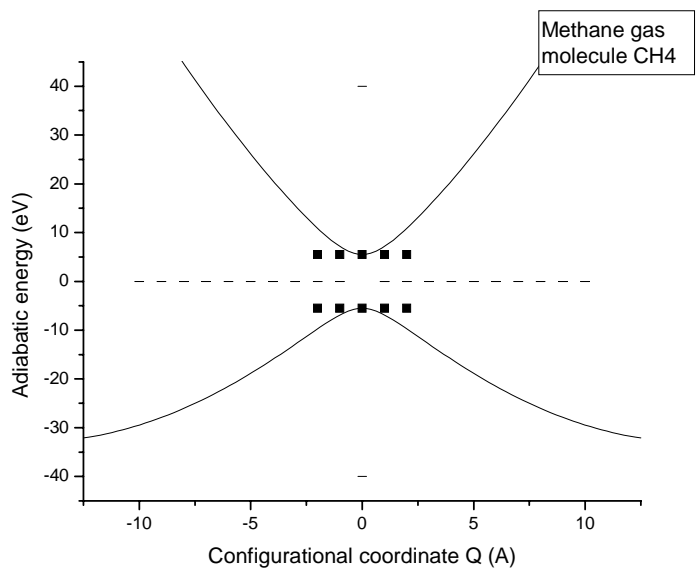

Figure 3

Adiabatic potentials through Eq. (12) for the $CH_4$ methane gas molecule (solid lines). The uncoupled TLS are shown by filled squares. The parameters were $K = 0.29$ eV/Å$^2$, $\Delta\varepsilon = 11$ eV, $G = 4.36$ eV/Å. Unlike the ammonia case in Figure 2, the high barrier ($\varepsilon_B \sim 0.27$ eV) keeps the methane molecule frozen in one of the lateral wells rendering it with a permanent electric dipole moment.

## 7. Brief survey of ammonia related work

### 7.1. Virtual design of enhanced-polarizability molecules

Although the ammonia molecule does not seem to provide the ideal enhanced-polarizability species, it is, nevertheless, interesting as a sample double-well character illustrating the basic requirements for the right unit. We summarize below published literature on the physical properties of the sample molecule. The papers are grouped into thirteen categories according to their main item. Although there may be a considerable overlap between different categories, we hope the classification will serve its purpose.

The following few papers may prove essential for designing enhanced-polarizability molecules:

The pseudo Jahn-Teller (PJT) effect as the only source of instability of molecular high-symmetry configurations in nondegenerate states has been investigated by means of a method of *ab initio* evaluation of the vibronic $K_\varepsilon$ and nonvibronic $K_0$ contributions to the curvature $K = K_0 + K_\varepsilon$ of APES in the distortion direction [16]. This method allows one to reveal the PJT origin of the instability along the distortion which overcomes the positive $K_0$. Further insight into the origin of the instability is reached by estimating the relative contribution of the most active excited states for different distortions. Numerical calculations are carried out on several series of molecular systems: planar $AH_4$, A = B$^-$, C, N$^+$, O$^{2+}$, Si, planar $BH_3$, $CH_3^-$, $NH_3$, octahedral $MH_6$, M = Sc$^{3-}$, Ti$^{2-}$, V$^-$, Cr, Mn$^+$, and octahedral $TeF_6^{2-}$ $IF_6^-$ and $XeF_6$.

The steric "lone pair" effect is well known, though less is known on whether a distortion of the high symmetry structure actually occurs and to what extent. DFT calculations have analyzed the energetic, steric and bonding properties of molecules $AX_3$ (A=N to Bi; X = H, F to I) [17,18]. The "lone pair" in the initial $D_{3h}$ geometry is of a central atom $pz$ character for the $NX_3$ and $AH_3$ molecules, while it is of a $s$ symmetry in all other cases with a strong delocalization toward the ligands. The stabilization of the distorted $C_{3v}$ geometry is mainly due to covalency effects, whereas steric interaction forces do not seem to play any significant role. Applying the conventional vibronic PJTE approach to the $D_{3h} \rightarrow C_{3v}$ transition [$A_1' \otimes (\alpha_2''+ \alpha_1') \otimes A_2''$ interaction], is a means to predict the extent of distortion and the corresponding energy gain. The vibronic coupling constant and the stabilization energy which determine the total $D_{3h} \rightarrow C_{3v}$ energy gain vary according to F > H > Cl > Br > I (A=N to Bi), and N > P > As > Sb > Bi (X=H,F), the dependence on A being only small or not present (X = Cl to I). Apparently, the hardest molecules are most susceptible to vibronic coupling, the vibronic coupling energy being approximately imaged by the hardness difference $\eta(C_{3v}) - \eta(D_{3h})$. The inverse trend is observed for the extent of angular distortion $\tau_\alpha$ from $D_{3h}$ to $C_{3v}$ symmetry; here the softest molecules such as $Sb(Bi)Br_3$ exhibit the largest deviations and $NH_3$ exhibits the smallest deviations from a $D_{3h}$ geometry. These trends of $\tau_\alpha$ are due to the strong influence of the force constant, which represents the $C_{3v} \rightarrow D_{3h}$ restoring energy. The electronegativity difference did not characterize the lone pair effect.

The underlying papers deal with properties of ammonia and methane molecules which may be useful for understanding the physics of the 'enhanced-polarizability molecule'.

### 7.2. Excitation spectra

The K-shell excitation spectra of water, ammonia, and methane have been measured in photoabsorption experiments using synchrotron radiation and high-resolution monochromator [19]. The excitation energies and relative intensities compare well with values calculated to second-order approximation for the polarization propagator. To determine the extent of admixing of valence excitations to the Rydberg manifolds, the X-H bond lengths have been varied. For $H_2O$, the two lowest-energy bands are due to the O 1s→$4a_1$/3s and O 1s→$2b_2$/3p transitions and have a strong valence character; their width indicating that both excitations are dissociative. The $NH_3$ and $ND_3$ spectra are also broad, due to possible dissociation, unresolved vibrational fine structure ($\nu_2$ mode), and to a Jahn-Teller instability. The valence character is concentrated in the lowest excited state of the Rydberg ns manifold but is distributed more uniformly over the np(e) manifold. The weak dipole-forbidden C 1s→3s ($a_1$) transition in $CH_4$ and $CD_4$ is accompanied by a vibrational structure due to the $\nu_4$ mode, indicating that it derives its intensity from the vibronic coupling to the C 1s→3p ($t_2$) transition. The structure of the latter band is due to Jahn-Teller coupling. The higher Rydberg np excitations contain considerable valence character.

### 7.3. Inversion barriers

Inversion barriers for the group-15 hydrides $NH_3$, $PH_3$, $AsH_3$, $SbH_3$ and $BiH_3$ have been studied using *ab initio* SCF methods including electron correlation and relativistic effects [20]. A modified symmetry inversion potential is introduced to describe the conversion from the minimum $C_{3v}$ arrangement through the $D_{3h}$ transition state. Tunneling rates and frequencies are calculated at the HF and MP level within the BWK approximation  At the MP level, the calculated $0^+/0^-$ $\nu_2$ frequency splitting of the vibronic ground state of $NH_3/ND_3$ (0.729 cm$^{-1}$/0.041 cm$^{-1}$) agrees with the experimental values (0.794 cm$^{-1}$/0.053 cm$^{-1}$). Correlation effects have not changed the barriers significantly but relativistic effects increased the $BiH_3$ barrier by 81.6 kJ/mol. Nonrelativistic and relativistic EH calculations suggest that the $a_1$ orbital, antibonding to Bi 6s, relieves part of its antibonding character near the equilateral geometry, due to the relativistic radial contraction of the 6s orbital hence increasing the barrier. In the planar transition state this orbital is a nonbonding $a_2''$. The increasing barrier heights from $NH_3$ to $BiH_3$ are explained by a second-order JT distortion of the trigonal planar geometry.

### 7.4. Electronic structure

The electron spectra of the $2A_1 \leftarrow 1A_1$ and the $2E \leftarrow 1A_1$ transitions of $NH_3$ have been measured with resolution in the search of a fine structure [21]. The 1st transition, $2A_1 \leftarrow 1A_1$, contains 2 vibrational progressions which are assigned to the $\nu_2$ bending mode. The 2nd transition, $2E \leftarrow 1A_1$, consists of 2 overlapping electron bands due to Jahn-Teller splitting of the 2E state.

Using monochromatic vacuum-UV radiation generated by two-photon resonant sum frequency mixing the electronic transitions of $NH_3$ have been investigated in the 67,000-73,000 cm$^{-1}$ region by detecting the fluorescence from predissociation fragments of $NH_3$ [22]. Vibronic bands of the $B_1E''$ (29, 210, 211, 212, 2731, 2831) state are clearly identified and relevant molecular constants deduced. The predissociation in the ~$B_1E''$ state is nearly independent of the rotational quantum numbers (J,K) within one vibrational band.

The electronic spectrum and the absolute photoabsorption oscillator strengths for the valence shell of $NH_3$ have been measured using high-resolution (0.048 eV fwhm) and low-resolution (~1 eV fwhm) dipole (e,e) spectroscopy in the photon energy ranges 5.0-31 and 5.5-200 eV, respectively [23]. The high-resolution data obtained at vibrational resolution are used to determine the integrated oscillator strengths of several vibronic transitions below the first ionization potential.

### 7.5. Vibronic structure

Several new vibronic levels of the ~B1E'' state of $NH_3$ have been found using IR-optical double resonance and supersonic jet spectroscopy [24]. Rotational analyses of these bands are made. A small but significant Jahn-Teller distortion is present (less than the zero point motion). Vibronic interactions with nearby electronic states are necessary for explaining the observed vibronic energy levels.

The vibrational structure of photoelectron spectra of PJTE molecules are discussed in [25]. On account of the vibronic interaction, the proportionality of the line intensities violates the Franck-Condon factor and the band formed becomes dependent on the angle of escaping photoelectrons.

In real-time two-color pump-probe ionization experiments of $NH_3$ and $ND_3$ using femtosecond 155 nm excitation pulses the lifetimes of several vibronic B levels have been measured [26]. The lifetimes obtained are discussed in terms of the nonadiabatic coupling between B and A vibronic levels.

### 7.6. Rotational structure

The rotational levels of an electronically degenerate planar $XY_3$ molecule have been derived [27]. There is a strong l-type doubling in levels with no vibrational angular momentum; it has 2 main causes, namely, a rotational-electronic interaction and a rotational Jahn-Teller interaction. The l-type doubling of other levels is largely quenched by the vibronic coupling. A mechanism is suggested to account for the variation of the l-type doubling with the vibronic state.

Using microwave-detected microwave-optical double resonance, the homogeneous linewidths of individual rovibrational transitions in the ~A state of $NH_3$, $NH_2D$, $NHD_2$, and $ND_3$ [28] have been measured. The excited state spectroscopic data are used for characterizing the height of the dissociation barrier and the mechanisms by which the molecule employs its excess vibrational and rotational energies to overcome this barrier. To interpret the vibronic widths, a 1D local mode potential is developed along a N-H(D) bond. However, the calculations suggest a barrier height of roughly 2100 cm$^{-1}$ well below the *ab initio* prediction. The rotational enhancement of the predissociation rates in the $NH_3$ 21 level is dominated by Coriolis coupling while centrifugal effects dominate in $ND_3$.

### 7.7. Rydberg states

The JTE in the pE" Rydberg series of $NH_3$ is investigated within a simple multi-channel-quantum-defect-type model [29]. The parameters of the model are determined from observed IR-optical double resonance spectra of the ~$B_1E$" ($3p_{x,y}$) band system. JTE induced vibrational autoionization effects and pronounced perturbations of high Rydberg members are analyzed.

### 7.8. Predissociation

The spectroscopy and predissociation of the vibronic levels of the ~$A_1A$"$_2$ excited states of $NH_3$ and $ND_3$ are studied via ~A-~X dispersed emission spectra, ~C'-~A dispersed emission and ~C'-~A stimulated emission pumping, following two-photon excitation to a selected intermediate level [30The lowest levels predissociate by H(D) atom quantum tunneling, the higher levels by vibrational rearrangement.

### 7.9. Photodissociation

Vibrationally stimulated photodissociation combined with Doppler spectroscopy and time-of-flight detection of H-atoms provides information on the photofragmentation dynamics from rovibrational states of ~$A_1A_2$"-state ammonia [31]. The competition between adiabatic dissociation forming excited-state $NH_2(2A_1)$ + H and nonadiabatic dissociation leading to ground-state $NH_2(2B_1)$ + H products changes drastically for dissociation from different parent levels prepared by double-resonance excitation. The H-atom speed distributions suggest that the nonadiabatic dissociation channel is the major pathway except for dissociation from the antisymmetric N-H stretching (31) parent level, which forms exclusively $NH_2(2A_1)$ + H.

Vibrationally mediated photodissociation combined with Rydberg H atom photofragment translational spectroscopy reveals the state-to-state photofragmentation dynamics of selected parent rovibronic levels of ~$A_1A_2$" state ammonia [32]. Analysis of the time-of-flight spectra determines the population of quantum states in the NH$_2$ partner fragment for dissociation from the excited state bending vibration (41). Dissociation from the bending state produces rotationally excited NH$_2$ predominantly (~75%) in its ground vibrational state.

The photodissociation dynamics of ~A state ammonia molecules has been investigated for D-substituted derivatives NH$_2$D, NHD$_2$, and ND$_3$ measuring the homogeneous linewidths of action spectra [33]. The spectral linewidth decreases in proportion to the number of D atoms. This shows that the dissociation. rate in the ~A state is additive.

The dissociation dynamics for NH$_3$(~A) → H + NH$_2$(~X) has been calculated using time-dependent quantum theory and an *ab initio* potential in the three dimensions of NH stretching, the out-of-plane motion and HNH bending [34]. The calculations use a Hamiltonian combining internal coordinates for NH$_2$ and Jacobi coordinates for the relative motion of the H atom. The calculated dissociation rates are inferior to observed ones indicating that the effective potential barrier to dissociation may be too high. Excited state vibronic resonant energies, resulting product-state population distributions and state-dependent product recoil anisotropy compare favorably with experiment,. Dissociation through the 00 and 21 levels of the ~A state proceeds by quantum tunneling through the barrier. Scattering of the outgoing waves at a conical intersection of the ~A and ~X adiabatic surfaces then accelerates the out-of-plane motion to generate high rotation of NH$_2$ about its inertial axis. Forces at this intersection also generate excitation of NH$_2$ bending. Higher level dissociation is mediated by intramolecular vibrational energy redistribution.

## 7.10. Photofragmentation

A spectral cross-correlation method is extended to analyze product-state dynamical data from photofragmentation [35]. Fragment product state vibrational distributions for the photodissociation of ammonia and deuterated ammonia are examined. The photodissociation dynamics of all four parent species (NH$_3$, NH$_2$D, ND$_2$H and ND$_3$) are studied simultaneously at 193.3 nm. The electronic emission spectra from the NH$_2$(A~ $^2A_1$), ND$_2$(A~ $^2A_1$), and NHD(A~ $^2A_1$) fragments are recorded by time-resolved Fourier transform IR spectroscopy.

The photofragmentation dynamics of NH$_3$ molecules following pulsed laser excitation to the 2 lowest levels (v'$_2$ = 0 and 1) of their ~A$_1$A$_2$'' excited state has been investigated by monitoring the time-of-flight spectra of the nascent H-atom products [36]. These spectra show well resolved structure whose analysis reveals that the majority of accompanying NH$_2$ (~X$_2$B$_1$) fragments are formed vibrationally unexcited, but with high levels of rotational excitation. The detailed energy disposal is sensitive to the initially excited parent vibronic (and even rovibronic) level.

## 7.11. Molecular ions

Adiabatic potentials as in Figure 2 have been reported using Eq. (12) for H$_3^+$ linear molecule ion (*G = 3.02* eV/Å, *K = 0,80* eV/Å$^2$, *Δε* = 12 eV) [8,15]. Accordingly, the dipolar double-well instability has proved inherent to the ground electronic state, while the excited electronic state has proved single-well.

### 7.11.1. Ammonia cation NH$_3^+$

The spectroscopic and dynamic aspects of JTE and PJTE interactions in the $NH_3^+$ cation have been studied within an *ab initio* based vibronic-coupling model [37]. Multireference 2nd-order perturbation theory is employed to obtain the potential energies of the ground state and the 1st excited state of $NH_3^+$ as function of symmetry-coordinate displacements. Vibronic-coupling parameters determining the FC, JTE, and PJTE activity of the normal modes are obtained from the *ab initio* data. The vibronic structures of the ~X $2A_1$ and ~A $2E$ photoelectron bands of $NH_3$ are calculated by numerical diagonalization of the vibronic Hamiltonian matrix. All 6 vibrational degrees of freedom are taken into account. The effects of JTE and PJTE interactions on the band shape of the ~A $2E$ photoelectron band are analyzed.

### 7.11.2. Amidogen radicals $NH_2$

The observation of two high energy vibronic transitions in the $NH_2$ amidogen radical, are reported [38]. The $NH_2$ radical has been produced in a laser photolysis experiment and expanded in a molecular beam machine. Laser excitation and dispersed fluorescence spectra provided term values and rotational constants in good agreement with predictions.

The adiabatic dissociation dynamics of $NH_2D$(~A) and $ND_2H$(~A) has been probed by time-resolved Fourier transform IR emission spectroscopy [39]. A product-state spectral pattern recognition technique is employed to separate out the emission features arising from the different photofragmentation channels following the simultaneous excitation of mixtures of four parent molecules $NH_3$, $NH_2D$, $ND_2H$, and $ND_3$ at 193.3 nm. This excitation reflects the competition between two distinct dissociation mechanisms that sample two different geometries during the bond cleavage. A larger quantum yield for producing $ND_2$ (~A, $\varepsilon_2' = 0$) from the photodissociation of $ND_2H$ rather than $ND_3$ is attributed to the lower dissociation energy of the N-H as compared with the N-D bond and to the enhanced tunneling efficiency of H atoms over D atoms through the dissociation barrier to. Similarly, the quantum yield for producing the $NH_2$ (~A, $\varepsilon_2' = 0$) fragment is lower when an N-D bond must be cleaved in comparison to an N-H bond. Photodissociation of $ND_2H$ by cleavage of an N-H bond leads to an $ND_2$ (~A) fragment. The quantum yield for producing NHD (~A) is larger for cleavage of an N-H bond from $NH_2D$ than by cleavage of an N-D from $ND_2H$.

### 7.12. $NH_3$ clusters

Ultrashort 155 nm pulses have been used to study the dynamics of ammonia clusters excited to vibronic levels of the ~B and ~C' state [40]. For the monomer rather long lifetimes of about 8 ps for $NH_3$ and 65 ps for $ND_3$ are found while the clusters decay in the sub-ps region. In contrast to ammonia clusters excited to the ~A electronic state, for $(ND_3)n$ the lifetime of the one-photon excited vibronic levels of the ~B (and ~C') state decreases with the cluster size.

### 8. Discussion

The foregoing theory of VdW interactions between TLS, whether uncoupled or coupled vibronically, is equivalent to a second-order perturbation approach to the binding energy in which the role of a small parameter is played by the dipole-dipole coupling term of magnitude $p^2/R^3$ [7]. This imposes a stringent condition on the intermolecular bond lengths of the form $R \gg p^{2/3}$. As p is increased, e.g. in a Rydberg atom, so does $R$ if it is to meet the second-order criterion and the VdW structure should become more dilute. We regard this criterion as one imposed by the 2nd-order theory and assume that the general implications of the enhanced-polarizability premises for a colossal binding go far beyond.

We have already commented *in situ* on the colossal effect through dipole elevation on exciting a TLS, whether uncoupled or vibronically coupled, to a higher-lying Rydberg state. There being seemingly no interference whatsoever, the Rydberg elevation and the vibronic enhancement could go simultaneously in a coupled TLS.

We should also point out that Eq. (5) of the VdW cohesive energy is based on the additivity assumption which may not be true strictly but it rather comes as a convenient approximation.

Yet, a key point of the present discussion is the conditions that make the vibronic polarizability of a double-well character molecule effective experimentally. Indeed, the stringent requirement is that there should be at least one, possibly two, vibronic energy levels under the barrier to give rise to the gap narrowing and thereby to the enhanced polarizability. The rigorous sole-level condition reads $\varepsilon_B = \varepsilon_{JT}[1 - (\Delta\varepsilon / 4\varepsilon_{JT})]^2 \gg ½ \hbar\omega$. We remind that for a low gap 'small-polaron' material $\Delta\varepsilon \ll 4\varepsilon_{JT}$. Here $\varepsilon_{JT}$ is the electron–vibrational mode coupling energy.

Given the appropriate adiabatic potential energy surface (APES), the subsequent condition to be met in soft condensed matter is that the well-interchange frequency $\Delta\nu = ½ \Delta\varepsilon / h$ ($\Delta\varepsilon$ is the ground state vibronic gap or tunneling splitting) be largely superior to the reciprocal traversal time for the encounter of two on-coming molecules (no retardation). The latter may be estimated as the ratio of the average thermal velocity $v_T = (3k_BT/M)^{1/2}$ to the average separation $\mathcal{R} = (6/\pi N)^{1/3}$ in an uniform distribution of molecules with concentration $N$ and mass $M$. We get $\Delta\varepsilon_{vib} \gg 2hv_T / \mathcal{R}$. This sets a lower limit to the reduced interlevel gap energy which should not fall beyond in soft condensed matter. In hard condensed matter the encounter time is always too large, due to the very low molecular velocities, so the no retardation condition is easily met. For soft condensed matter or a gas state we set $M = 17$ a.u. (NH$_3$) to estimate $v_T = 6.6\times10^4$ cm/s. Now, we get $\Delta\varepsilon_{vib} \gg 4.4$ meV for $N = 10^{21}$ cm$^{-3}$ (dilute gas) corresponding to the average intermolecular separation of $\mathcal{R} = 12.4$ Å.

We see that the estimate in Section 5.1 based on the double-well character of the NH$_3$ ground state falls considerably below the no retardation lower bound on the reduced vibronic gap energy $\Delta\varepsilon$. From the adiabatic potentials shown in Figure 2 we get $\varepsilon_B = 0.53$ eV at $\varepsilon_{JT} = 5.16$ eV which hosts three sub-barrier levels at $\hbar\omega = 0.18$ eV. We also obtain $\Delta\varepsilon /4\varepsilon_{JT} = 0.678$ at $\Delta\varepsilon = 14$ eV which indicates that the small-polaron criterion $\Delta\varepsilon /4\varepsilon_{JT} \ll 1$ is not met strictly. In as much as the vibronic splitting increases in magnitude as the level energy is increased, it is possible that the no retardation condition may be met at the higher excited vibronic sub-barrier levels for the adiabatic potential energy profile shown in Figure 2.

Another crucial point is the temperature dependence of the vibronic polarizability, all the foregoing considerations being based on the low-temperature polarizability instead. Using Boltzmann statistics and the inequalities $p_0F \ll \Delta\varepsilon$ and $p_0F \ll k_BT$, where $F$ is the field and $p_0$ the ultimate vibronic dipole, the temperature-dependent (semiclassical) polarizability of a TLS in 1D has been derived to be [8]:

$$\alpha(T) = \alpha_0 \tanh(\Delta\varepsilon_{vib} / k_BT) \qquad (13)$$

Here $\alpha_0 = p_0^2/3\Delta\varepsilon_{vib}$ is the zero-point vibronic polarizability,. At high temperatures $k_BT \gg \Delta\varepsilon$, we easily get $\alpha(\infty) = p_0^2/3k_BT$, which is the reorientational polarizability of a system of dipolar molecules (decoupled TLS). We conclude that the vibronic effects are viable at sufficiently low temperatures only depending on the magnitude of the vibronic energy splitting $\Delta\varepsilon_{vib}$. Expressions for $\alpha(T)$ of more complex systems are also available, as can be found in reference [8] and cited references therein.

To summarize, the following are the inequalities that should be met in order to make the colossal binding effective in real small-polaron-like systems:

$\varepsilon_B = \varepsilon_{JT}[1 - (\Delta\varepsilon/4\varepsilon_{JT})]^2 \gg \hbar\omega$ (sub-barrier levels)
$\Delta\varepsilon_{vib} \gg h\langle v_T\rangle/\langle \Re\rangle$ (no retardation condition)
$p_0F \ll k_BT$ (linear field effect)
$\Delta\varepsilon_{vib} \gg k_BT$ (low temperatures).

Of these, (ii) is perhaps the most stringent. For this reason the colossal VdW binding may actually prove a rare occurrence in soft condensed matter, while being a more likely happening in hard condensed matter, due to negligible molecular velocities therein. Ammonia makes no exception in developing a colossal VdW binding in ground state and, possibly, in excited state within denser clouds spread near the earth's surface [3]. This could have a profound effect on the formation of ball lightning matter, one of the remaining enigmatic problems for present-day science [2,3].

## 9. Conclusion

We have presented for the first time a simple theory for the enhanced polarizability of vibronically coupled TLS which may give rise to colossal VdW binding of molecules in soft and hard condensed matter. Earlier, colossal VdW binding has been considered with regard to polaronic-exciton and vibronic-polaron systems [1,10]. At least a few additional conditions have to be met as well in order to make the theoretical predictions realized in real systems, among them not too low inversion barriers to secure sub-barrier levels, not too high barriers to secure high well-interchange frequencies, and sufficiently low temperatures at which the vibronic polarizabilities are meaningful at all. The combined necessity for the observation of all these requirements may turn the colossal binding into a rare occurrence in soft condensed matter.

Retardation effects have not been comprehensively taken into account presently but may be dealt with in further investigations. Another plan for the future is studying the nonlinear effects arising in strong fields which violate the $p_0F \ll k_BT$ condition.

Acknowledgement. We dedicate this paper to the memory of Professor S.G. Christov for his pioneering contributions to the field of electron-phonon interactions from which we have benefited throughout this work and many other related studies.

## Appendix 1

### Vibronic Coupling at TLS

The TLS Hamiltonian assuming the vibronic mixing of two static electronic states $|1\rangle$ and $|2\rangle$ by a single vibrational mode coordinate Q reads:

$$H = \tfrac{1}{2}[\mathbf{P}^2/M + KQ^2] + GQ\,[|2\rangle\langle1| + |1\rangle\langle2|] + \tfrac{1}{2}\Delta\varepsilon[|2\rangle\langle2| - |1\rangle\langle1|] \qquad (A1.1)$$

Following the adiabatic theorem to separate the electronic coordinates **r** from the nuclear coordinates Q, we define an adiabatic Hamiltonian:

$$H_{AD} = \tfrac{1}{2}KQ^2 + GQ\,[|2\rangle\langle1| + |1\rangle\langle2|] + \Delta\varepsilon[|2\rangle\langle2| - |1\rangle\langle1|] \qquad (A1.2)$$

The next adiabatic step is solving Schrodinger's equation $H_{AD}\Psi = E_{AD}\Psi$ by means of a linear combination in the $\{|1\rangle, |2\rangle\}$ basis with $Q$-dependent amplitudes; this implies that *r* and $Q$ are separated. Solving to first order, we get a secular equation with two roots of an adiabatic potential energy surface (APES):

$$E_{AD}(Q) = \tfrac{1}{2}KQ^2 \pm \tfrac{1}{2}\sqrt{[4G^2Q^2 + (\Delta\varepsilon)^2]} \qquad (A1.3)$$

Here $\Delta\varepsilon = |\varepsilon_2 - \varepsilon_1|$ is the interlevel energy gap. We also introduce Jahn-Teller's energy $\varepsilon_{JT} = G^2/2K$ to characterize the vibronic coupling. At weak coupling $\Delta\varepsilon \geq 4\varepsilon_{JT}$ the profile depicted by (A1.3) is a pair of single-well anharmonic parabolas centered at the origin. At strong coupling $\Delta\varepsilon < 4\varepsilon_{JT}$ the upper parabola (+) retains its general character, while the lower parabola (−) turns double well with the interwell barrier centered at the origin. The lateral wells are centered at $\pm Q_0$ with

$$Q_0 = \sqrt{(2\varepsilon_{JT}/K)}\,\sqrt{[1 - (\Delta\varepsilon/4\varepsilon_{JT})^2]} \qquad (A1.4)$$

the magnitude of the lateral displacement. The quantized vibronic energy levels in the lateral wells can be found by the standard procedure. Of particular interest for this study is the ground state vibronic energy splitting. We form the symmetric and antisymmetric combinations of the ground state wavefunctions pertaining to the lateral wells. Using harmonic-oscillator wave functions, the energy difference or gap of the split-off vibronic energy levels will be found to be:

$$\Delta\varepsilon_{vib} = \Delta\varepsilon\,\exp(-2\varepsilon_{JT}/\hbar\omega) \qquad (A1.5)$$

in the strong-coupling limit $4\varepsilon_{JT} \gg \Delta\varepsilon$. The exponential in (A1.5) is Holstein's reduction factor. The reduced gap $\Delta\varepsilon_{vib}$ determines the polarizability of the vibronic TLS. The origin of Holstein's reduction is in the small though finite overlap of ground-state wave functions of displaced oscillators.

## References

[1] M. Georgiev and J. Singh, Pairing of vibronic small excitons due to enhanced polarizability in a crystalline polarizable nonmetallic medium, *Phys. Rev. B* **1998**, *58*, 15595-15602.
[2] J.J. Gilman, Cohesion in ball lightning, *Appl. Phys. Lett.* **2003**, *83*, 2283-2284.
[3] M. Georgiev and J. Singh, Exciton matter sustained by colossal dispersive interactions due to enhanced polarizability: Possible clue to ball lightning, *physics / 0509212*.
[4] See the discussion on cohesion in ball lightning, Physicsweb, September 2003.
[5] M. Born and K. Huang, *Dynamical Theory of Crystal Lattices*, Clarendon, Oxford, 1988.
[6] D.P. Craig and T. Thirunamachandran, *Molecular Electrodynamics*, Academic, London, 1984.
[7] Yu.S. Barash, *Sily Van-der-Vaalsa*, Moskva, Nauka, 1988 (in Russian).

[8] I.B.Bersuker, The Jahn-Teller Effect and Vibronic Interactions in Modern Chemistry, Academic, New York, 1984. Russian translation: Effekt Yana-Tellera i vibronnye vzaimodeistviya v sovremennoi *khimii* (Moskva, Nauka, 1987).

[9] J. Jäckle, On the ultrasonic attenuation in glasses at low temperatures, *Z. Physik* **1972**, *257*, 212.

[10] M. Georgiev and M. Borissov, Vibronic pairing models for high-Tc superconductors, *Phys. Rev. B* **1989**, *39*, 11624-11632.

[11] T. Holstein, Studies of polaron motion. Part I. The molecular crystal model, *Ann. Phys. (N.Y.)* **1959,** *8*, 325-342. Part II. The "small" polaron, *Ann. Phys. (N.Y.)* **1959**, *8*, 343-389.

[12] A. Sommerfeld, Atombau und Spektrallinien, Bd. I & II, Vieweg, Braunschweig, 1951.

[13] I.B. Bersuker, Electronic Structure and Properties of Transition Metal Compounds. Introduction to the Theory (Wiley, New York, 1996). Russian original: I.B.Bersuker, Elektronnoe stroenie i svoystva koordinatsionnykh soedinenij, Leningrad, Khimiya, 1986.

[14] I.B. Bersuker and V.Z. Polinger, *Vibronic Interactions in Molecules and Crystals,* Springer, Berlin, 1991.

[15] I.B. Bersuker, N.N. Gorinchoi, V.Z. Polinger, On the origin of dynamic instability of molecular systems, *Theoretica Chimica Acta* **1984,** *66*, 161-172.

[16] I.B. Bersuker, N.B Balabanov, D. Pekker, J.E. Boggs, Pseudo Jahn-Teller origin of instability of molecular high-symmetry configurations: Novel numerical method and results, *J. Chem. Phys*. **2002,** *117*,10478-10486.

[17] D. Reinen, M. Atanasov, DFT calculations of the "lone pair" effect - a tool for the chemist to predict molecular distortions? NATO Science Series, II: Mathematics, Physics and Chemistry (Vibronic Interactions: Jahn-Teller Effect in Crystals and Molecules) *39*, Kluwer Academic Publishers, Dordrecht, 2001, p.p. 83-95.

[18] M. Atanasov, D. Reinen, Density Functional Studies on the Lone Pair Effect of the Trivalent Group (V) Elements: I. Electronic Structure, Vibronic Coupling, and Chemical Criteria for the Occurrence of Lone Pair Distortions in $AX_3$ Molecules (A=N to Bi; X=H and F to I), *J. Phys. Chem. A* **2001,** *105*, 5450-5467.

[19] J. Schirmer; A.B. Trofimov, K.J. Randall, J. Feldhaus, A.M. Bradshaw, Y. Ma, C.T. Chen, F. Sette, K-shell excitation of the water, ammonia, and methane molecules using high-resolution photoabsorption spectroscopy, *Phys. Rev. A* **1993**, *47*, 1136-1137.

[20] P. Schwerdtfeger, L.J. Laakkonen, P. Pyykko, Trends in inversion barriers. I. Group VA element hydrides, *J. Chem. Phys*. **1992**, *96*, 6807-6819.

[21] J.W. Rabalais, L. Karlsson, L.O. Werme, T. Bergmark, K. Siegbahn, Analysis of vibrational structure and Jahn-Teller effects in the electron spectrum of Ammonia, *J. Chem. Phys.***1973**, *58*, 3370-3372.

[22] Li Xinghua, C.R. Vidal, Predissociation supported high-resolution vacuum ultraviolet absorption spectroscopy of excited electronic states of $NH_3$, *J. Chem. Phys*. **1994**, *101*, 5523-5528.

[23] G.R. Burton, Wing Fat Chan, G. Cooper, C.E. Brion, The electronic absorption spectrum of ammonia in the valence shell discrete and continuum regions. Absolute oscillator strengths for photoabsorption (5-200 eV), Chem. Phys. **1993**, *177*, 217-231.

[24] J.M. Allen, M.N.R. Ashfold, R.J. Stickland, C.M. Western, The ~B1E" state of ammonia: the Jahn-Teller effect revealed by infrared-optical double resonance, *Molecul. Phys*. **1991**,*74*, 49-60

[25] I.Ya. Ogurtsov, Angular dependence of the photoelectron spectra of molecules having the pseudo-Jahn–Teller effect, *Zh. Strukturnoi Khimii* **1986**, 27**,** 65-68.

[26] H.-H. Ritze, W. Radloff, I.V. Hertel, Decay of the ammonia B state due to nonadiabatic coupling, *Chem. Phys. Letters* **1998**, *289*, 46-52.

[27] M.S. Child, H.L. Strauss, Causes of l-type doubling in the 3p(.EPSILON.") Rydberg state of ammonia, *J. Chem. Phys.* **1965**, *42*, 2283-2292.

[28] S.A. Henck, M.A. Mason, Wen-Bin Yan, K.K. Lehmann, S.L. Coy, Microwave detected, microwave-optical double resonance of $NH_3$, $NH_2D$, $NHD_2$, and $ND_3$. II. Predissociation dynamics of the ~A state, *J. Chem. Phys.* **1995**, *102*, 4783-4792.

[29] A. Staib, W. Domcke, Vibronic coupling in the pE" Rydberg series of $NH_3$, *Chem. Phys. Letters* **1993**, *204*, 505-510.

[30] M.N.R. Ashfold, C.L. Bennett, R.N. Dixon, Dissociation dynamics of ammonia (~$A_1A"_2$), *Faraday Discussions of the Chemical Society* **1986**, *82*, 163-175.

[31] A. Bach, J.M. Hutchison, R.J. Holiday, F.F. Crim, Competition between Adiabatic and Nonadiabatic Pathways in the Photodissociation of Vibrationally Excited Ammonia, *J. Phys. Chem. A* **2003**, *107*, 10490-10496.

[32] A. Bach, J.M. Hutchison, R.J. Holiday, F.F. Crim, Photodissociation of vibrationally excited ammonia: Rotational excitation in the $NH_2$ product, *J. Chem. Phys.* **2003**, *118*, 7144-7145.

[33] A. Nakajima, K. Fuke, K. Tsukamoto, Y. Yoshida, K. Kaya, Photodissociation dynamics of ammonia ($NH_3$, $NH_2D$, $NHD_2$, and $ND_3$): rovibronic absorption analysis of the A-~X transition, *J. Phys. Chem.* **1991**, *95*, 571-574.

[34] R.N. Dixon, Photodissociation dynamics of ~A state ammonia molecules. III. A three-dimensional time-dependent calculation using ab initio potential energy surfaces, *Molecular Phys.* **1996**, *88*, 949-977.

[35] J.P. Reid, R.A. Loomis, S.R. Leone, Characterization of dynamical product-state distributions by spectral extended cross-correlation: Vibrational dynamics in the photofragmentation of $NH_2D$ and $ND_2H$, *J. Chem. Phys.* **2000**, *112*, 3181-3191.

[36] J. Biesner, L. Schnieder, J. Schmeer, G. Ahlers, Xiaoxiang Xie, K.H. Welge, M.N.R. Ashfold, R.N. Dixon, State selective photodissociation dynamics of ~A state ammonia. I., *J. Chem. Phys.* **1988,** *88*, 3607-3616.

[37] C. Woywod, S. Scharfe, R. Krawczyk, W. Domcke, H. Koppel, Theoretical investigation of Jahn-Teller and pseudo-Jahn-Teller interactions in the ammonia cation, *J. Chem. Phys.* **2003,** *118***,** 5880-5893.

[38] J. Schleipen, J.J. Ter Meulen, L. Nemes, M. Vervloet, New vibronic states of amidogen ($NH_2$) observed in ammonia photolysis, *Chem. Phys. Letters* **1992,** *197*, 165-170.

[39] J.P.Reid, R.A. Loomis, S.R. Leone, Competition between N-H and N-D Bond Cleavage in the Photodissociation of $NH_2D$ and $ND_2H$, *J. Phys. Chem. A* **2000,** *104*, 10139-10149.

[40] Th. Freudenberg, V. Stert, W. Radloff, J. Ringling, J. Guedde, G. Korn, I.V. Hertel, Ultrafast dynamics of ammonia clusters excited by femtosecond VUV laser pulses, *Chem. Phys. Letters* **1997,** *269*, 523-529.